\documentclass[aps,twocolumn,pra,showpacs,floatfix]{revtex4}
\usepackage{epsfig}
\usepackage{dcolumn}

\begin{document}

\title{\bf Isotope shift and search for metastable superheavy elements in astrophysical data} 

\author{V. A. Dzuba$^1$,  V. V. Flambaum$^{1,2}$,  and J. K. Webb$^1$}
\affiliation{$^1$School of Physics, University of New South Wales,
Sydney 2052, Australia}
\affiliation{$^2$Helmholtz Institute Mainz, Johannes Gutenberg University, 55099 Mainz, Germany}

\date{\today}

\begin{abstract}
Spectral lines belonging to the short-lifetime heavy radioactive elements up to Es ($Z$=99) have been  found in the spectra
of the Przybylski's star. We suggest that these unstable elements  may be decay products of a "magic" metastable nucleus belonging to the the island of stability where the nuclei have a magic number of neutrons $N=184$. The laboratory-produced nuclei  have a significantly smaller number of neutrons. To identify  spectra of the $N=184$ isotopes of these  nuclei  and their neutron-reach  superheavy decay products in astrophysical data we calculate the isotope shift  which should be added to the laboratory - measured wavelenghs.  The results for the isotopic shifts in the strongest optical electromagnetic transitions  in  No, Lr, Nh, Fl,and $Z$=120 elements are presented. 
 \end{abstract} 
\maketitle

The existence of the hypothetical {\em island of stability} for superheavy elements (atoms with nuclear charge $Z \ge 104$)
is an important problem of modern nuclear physics.  All known nuclei with $Z>98$ 
have short lifetimes varying from fractions of seconds to hundreds of days. On the other hand, the theoretically estimated lifetime of the Fl nucleus with $Z=114$, $N=184$ is $10^7$ years ~\cite{Oganessian2004,Oganessian2013}. The nuclear shell model states that nuclei are most stable when both
protons and neutrons fully occupy closed shells (double magic nuclei). For superheavy elements with spherical nuclei, 
the magic neutron number is believed to be 184, while magic proton numbers are $Z=114$, $Z=120$ and
$Z=126$ (see, e.g. \cite{Oganessian2004,Oganessian2013}). 
(Magic numbers $Z=122$, $N=172,178,182,194$ are also mentioned in the 
literature~\cite{Oganessian2004,Oganessian2013,Patra2000,Santhosh17}.)
Therefore, the prospective candidates for the island of stability include the $^{298}_{114}$Fl, 
$^{304}_{120}$Ubn, $^{310}_{126}$Ubh and some other isotopes.

Superheavy elements are not found in nature but are produced in laboratories by colliding lighter atoms. All elements up to 
$Z=118$ have been synthesised so far ~\cite{Oganessian2013}. However, all superheavy elements synthesised in a laboratory
are neutron-poor elements, with the number of neutrons being significantly smaller than required to make the most stable 
isotopes. For example, the heaviest isotope of Fl, produced in Dubna ~\cite{Oganessian2004}, $^{292}_{114}$Fl, is six neutrons
short of the magic number $N=184$. This is a common problem because the Coulomb repulsion energy for protons increases as $Z^2$ and to compensate for this by the attractive strong interaction energy, the number of neutrons $N$ should increase with $Z$ faster than the number of protons. 
Therefore, the very large number of neutrons, $N=184$, needed 
for a more stable superheavy element, cannot be obtained by colliding any pair of lighter elements where the $N/Z$ ratio is smaller 
than that in the island of stabiltiy. 

There is an alternative way of searching for the island of stability by looking for
traces of
superheavy elements  in astrophysical data (see, e.g.~\cite{Polukhina2012}).  For example, optical lines of many actinide atoms and ions up to einsteinium
(Es, $Z=99$) have been found in the spectra 
of the Przybylski's star~\cite{Gopka2008}. Since these lines belong to short-living isotopes the natural question is 
how the isotopes were produced. One possible answer is that they were produced via decay of heavier long-living elements.

Heavy elements are dispersed into the interstellar medium during supernova explosions. The neutron flux during the supernova explosion is very high, and this may lead to the production of the $N=184$ and other neutron reach isotopes. 
If they are close to the island of stability they may have sufficiently long lifetime to survive to present time and
decay to isotopes of actinides and  other elements. 
One needs to know frequencies of the strong electric dipole transitions for superheavy elements
to search for them in astrophysical spectra. The heaviest element for which one such frequency is 
measured is No ($Z=102$)~\cite{Block2015a}. Work is underway for similar measurements
in Lr ($Z=103$)~\cite{sato2015measurement,Sato2015,Block2015}. There are good prospects for further progress in this field.

There are many high-quality atomic spectra calculations for superheavy elements (see, 
e.g.~\cite{E-E113,E114a,E115,eliav2015,borschevsky2013ab,DSS14,GD15,DD16}). However,
the accuracy of the calculations is not sufficiently high to reliably identify spectral lines in astrophysical data.  A possible solution involves the following three-stage process:
\begin{enumerate}
\item 
Measure the  frequencies of strong electric dipole transitions in a laboratory-produced superheavy element.  This will be a neutron-poor isotope.
\item To find the frequencies for the  more stable neutron-rich isotope, the isotopic shifts are calculated and added to the laboratory frequencies. 
\item The results are used to search for the spectral lines from the more stable neutron-rich isotope in 
astrophysical data.
\end{enumerate}

\begin{table*}
\caption{\label{t:is}
Isotope shift for strong electric dipole transitions from the ground state of some
heavy elements. The shift is given by $\delta E = a (A_1^{1/3} - A_2^{1/3})$,
where $A_1$ and $A_2$ are atomic numbers of the two isotopes,
the values of the parameter $a$  are presented in the last column.
 $A_0 = Z+184$ is the atomic number of the more stable isotope with the neutron number $N=184$, $A_s$ is the atomic 
 number of the heaviest synthesised  isotope.  
The "Frequency"  column presents experimental (E) or
theoretical (T) values for the frequency of the transition found in literature.
Theoretical uncertainty is presented in parentheses.}
\begin{ruledtabular}
\begin{tabular}{rlcc lcl ccr}
\multicolumn{2}{c}{Atom} & & &
\multicolumn{3}{c}{Transition}& 
\multicolumn{1}{c}{Frequency}  &  
\multicolumn{1}{c}{Reference} & 
\multicolumn{1}{c}{$a$}  \\
\multicolumn{1}{c}{$Z$} &
\multicolumn{1}{c}{Symbol} &
\multicolumn{1}{c}{$A_0$} &
\multicolumn{1}{c}{$A_s$} &
&& &\multicolumn{1}{c}{(cm$^{-1})$} && \multicolumn{1}{c}{(cm$^{-1})$} \\
\hline





102 & No & 286 & 259 & $7s^2 \ ^1$S$_0$ &-& $7s7p \ ^1$P$^o_1$ & 29961.457$^{+0.041}_{-0.007}$\footnotemark[1] & E~\cite{Block2015a} & 34 \\

103 & Lr & 287 & 266 & $7s^27p \ ^2$P$^o_{1/2}$ &-& $7s^28s \ ^2$S$_{1/2}$ & 20253(500) & T~\cite{DSS14} & -19 \\

113 & Nh & 297 & 286 & $7s^27p \ ^2$P$^o_{1/2}$  &-& $7s^28s \ ^2$S$_{1/2} $ &  36041(440) &T~\cite{DD16} &  -17 \\

114 & Fl & 298  & 292 & $7p^2 \ ^1$S$_0$ &-& $7p8s \ ^1$P$^o_1$ & 43876(310) & T~\cite{DD16} &  -2.6 \\

120 & Ubn & 304  &  & $8s^2 \ ^1$S$_0$ &-& $8s8p \ ^1$P$^o_1$ & 27559(200) & T~\cite{GD15} & 148 \\
\end{tabular}
\end{ruledtabular}
\footnotetext[1]{For the $^{254}_{102}$No isotope.}
\end{table*}

One needs to know the isotope shifts for superheavy elements to follow this path. 
The heaviest elements for which isotope shift experimental data are available are Pu, Am, Cm~\cite{Actinides}
and No~\cite{Block2015a}. These data can be used for searching heavier isotopes of the elements. 
However, no experimental data are available for the superheavy elements in the vicinity of  the island of stability, such as $_{114}$Fl, $_{120}$Ubn, etc. 

In the present work we calculate the isotope shift for some elements within nuclear charge range $102 \le Z \le 120$,
including candidates for the island of stability, Fl and Ubn. The isotope shift
in superheavy elements is strongly dominated  by the field (volume) shift~\cite{Sobelman}. Therefore, we ignore the mass
shift and calculate only the energy shift due to the change of the nuclear charge  radius. We assume a Fermi
distribution for the nuclear charge with the radius given by $R_N = 1.1 A^{1/3}$~fm with a skin 
thickness of 2.3~fm. 
Here $A$ is the number of nucleons in the nucleus.

We use a combination of the configuration interaction (CI) method with many-body perturbation
theory (MBPT) to perform the calculations (the CI+MBPT method \cite{DzuFlaKoz96,DzuJoh98,Dzu05}). The field shift
is obtained by repeating calculations with different values of nuclear radius and taking differences between the results. 
We consider only the strongest optical transitions from the ground state, which are the $s-p$ electric dipole transitions 
to states of opposite parity having the same total electron spin (the change of the spin leads to a partial suppression of 
the $E1$ transition probability since the electric dipole operator conserves the spin). 

Since $R_N \propto  A^{1/3}$ it is convenient to approximate the calculated isotope shift $\nu$ by an analytic formula: 
\begin{equation}
\nu = a(A_1^{1/3}-A_2^{1/3}).
\label{eq:nu}
\end{equation}
The coefficient $a$ in (\ref{eq:nu}) is found by fitting to the results of the CI+MBPT numerical calculations.
The values of $a$ for No, Lr, Nh, Fl and Ubn are presented in Table~\ref{t:is}.

For No there is an additional source of the information about the isotope shift. We may use the experimental value of the isotope shift:   
$\nu(^{252}{\rm No})-\nu(^{254}{\rm No}) = 0.32$~cm$^{-1}$~\cite{Block2015}.
Then  Eq. (\ref{eq:nu})  provides the value  $a=19$~cm$^{-1}$ for No.
This is about 1.8 times smaller than the calculated value $a=34$~cm$^{-1}$ (see Tavle~\ref{t:is}). 
The difference is partly due to an uncertainty of the atomic calculations and partly due to an 
uncertainty in the nuclear radius change. The formula $R_N \propto  A^{1/3}$ gives an average trend in the dependence of $R_N$ on $A$ 
while the actual change of the nuclear radius depends on what orbitals are occupied by the additional neutrons.  Thus, we may trust  the 
dependence of $R_N \propto  A^{1/3}$  only for a significant change in $A$ when the shell model fluctuations are relatively suppressed in 
comparison with the average trend.
The $^{254}$No isotope of nobelium is the heaviest element for which experimental frequency is available.
The isotope with the magic neutron number $N=184$ is $^{286}$No. The difference in neutron numbers for these two isotopes is
large, $\Delta N = 286-254=32$. Therefore, the $R_N \propto  A^{1/3}$ trend should hold to a high precision.
This means that prediction transition frequency for the $^{286}$No isotope using theoretical isotope shift might be more 
accurate than using experimental shift.
The value for the frequency is
\begin{eqnarray}
&&\nu(^{286}{\rm No}) =  29961.457 + \label{eq:Not} \\ 
&& 34(254^{1/3} - 286^{1/3}) = 29952.8 \ {\rm cm}^{-1}, \nonumber
\end{eqnarray}
if the theoretical value $a=34$~cm$^{-1}$ is used (and $\nu(^{286}{\rm No}) = 29956.6 \ {\rm cm}^{-1}$ 
if the experimental value $a=19$~cm$^{-1}$ is used).

It is important to have an independent way of estimating the isotope shift. We use analytic solutions for the isotope shift problem in the 
single-electron approximation.
 The accuracy of this approach is significantly lower than the accuracy of numerical many-body calculations.
However, it helps to avoid mistakes and it can be used to extrapolate the isotope shift from lighter atoms (where experimental data are
available) to heavier elements with a similar electron structure. Such an extrapolation makes sense since the relative value of the many-body corrections to the single-electron formula is approximately the same in atoms with  similar structure of external shells.

The energy shift of an $s$ state due to change of the nuclear radius is given by ~\cite{Sobelman,Smorodinsky} 
\begin{eqnarray}
&&\delta \epsilon_s = 8\sqrt{2}\xi K \epsilon_s^{3/2}\frac{\gamma+1}{\Gamma(2\gamma+1)^2(2\gamma+1)}  \label{eq:Rf} \\
&& \times (2ZR_N)^{2\gamma}\frac{\delta R_N}{R_N} {\rm Ry}, \nonumber \\
&&{\rm where} \ \ \xi = \frac{2\gamma^2(2-\gamma)(2\gamma+1)}{(\gamma+1)(2+\gamma)}. \label{eq:xi}
\end{eqnarray}
Here $\gamma = \sqrt{1-(\alpha Z)^2}$, $\Gamma$ is the gamma-function,
$R_N$ is nuclear radius, $\epsilon_s$ is the ionization energy of the $s$ orbital in atomic units (2 Ry), Ry=109737 cm$^{-1}$.   The parameter $\xi$ describes the non-perturbative relativistic correction which is important at $Z>110$ (when $Z$ approaches 137, the finite nuclear size effect is not a small perturbation because of the collapse of the point-like nucleus spectrum for $Z >137$). The analytic result in Eq. (\ref{eq:Rf}) with the factor $K=1$ was obtained in Ref.~\cite{Sobelman,Smorodinsky} for the unrealistic charge distribution where  all the charge is located on the surface of the sphere of the radius  $R_N$. A more realistic case with the electric charge homogeneously distributed inside the sphere may be obtained by introducing an approximate correction factor $K=3/(2 \gamma +3)$ to the isotope shift in Eq. (\ref{eq:Rf}) (this factor follows from the perturbative treatment of the isotope shift for the cases of the surface and volume charge distributions  - see Racah-Rosental-Breit formulae for the isotope shift presented in the book   ~\cite{Sobelman}). 
With this factor included the analytic results  are close to the experimental values of the isotope shifts (see below).

The formula for the isotopic shift for $p$ wave is not presented in the text books. Indeed,  in the non-relativistic approximation, 
the single-particle isotope shift for  $p$ orbitals is zero since the $p$ orbitals vanish at the origin $r=0$. However, the relativistic 
$p_{1/2}$ wave  density near the nucleus is proportional to the $s$ wave density. The relativistic calculation of the proportionality factor gives  the following isotopic shift for $p_{1/2}$ states in heavy atoms:
\begin{equation}
\delta \epsilon_p = \frac{1-\gamma}{1+\gamma}\left(\frac{\epsilon_p}{\epsilon_s}\right)^{3/2} \delta \epsilon_s.
\label{eq:dp}
\end{equation}
Thus, in the superheavy atoms, where $(1-\gamma) \sim 1$, the $p_{1/2}$ shift is not suppressed significantly.

Formulae (\ref{eq:Rf},\ref{eq:xi},\ref{eq:dp}) give $a=160$~cm$^{-1}$ for Ubn in good agreement with the calculated value $a=148$~cm$^{-1}$(see Table~\ref{t:is}).
The difference can be attributed to the many-body effects neglected in (\ref{eq:Rf},\ref{eq:xi},\ref{eq:dp}).
The same formulae give $a=9.3$~cm$^{-1}$ for Ra, which is a lighter analogue of Ebn. On the other hand, the fitting of the experimental data~\cite{Ra-IS-exp}
gives $a=12$~cm$^{-1}$. 

Note the very rapid increase of the isotope shift when the nuclear charge is approaching 120. 
Such an increase is clearly seen in both the analytic formulae and the numerical many-body calculations. 
This is the result of the relativistic effects mentioned above.

Another feature of the isotope shift is its smaller value in  $7p_{1/2}$ - $8s$ transitions  in Lr, Nh and Fl.
This is the result of the cancellation of the shifts of the $7p_{1/2}$  and  $8s$ states. For example, for Nh the formulae 
give nearly the same  shifts of the energies of the lower $7p_{1/2}$ and upper $8s$ states: $\delta \epsilon(7p_{1/2}) = 1.11$~cm$^{-1}$, 
while  $\delta \epsilon(8s_{1/2}) = 0.97$~cm$^{-1}$. 
When such a cancellation occurs, the accuracy of the analytic formulae is low:
the formulae give $a=-4.5$~cm$^{-1}$ for Nh in apparent disagreement with the result of the many body calculations $a=-17$~cm$^{-1}$ (see Table \ref{t:is}). 

Thus, the formulae
 (\ref{eq:Rf},\ref{eq:xi},\ref{eq:dp}) give a reasonable accuracy when an $s$ state is lower than a  $p$ state and indicate 
 cancellation when a $p_{1/2}$ state is lower than an  $s$ state.

We hope that this work provides a motivation for a further progress in
the measurements of the transition frequencies for superheavy elements, calculations of the isotope shifts and search 
for the corresponding transitions in astrophysical spectra.

We are very grateful to J.-T. Hu and  G. Gr$\ddot{\rm u}$ning for the information about measured heavy element spectra.
This work is supported by the Australian Research Council and Gutenberg Fellowship. 

\bibliographystyle{apsrev}


\end{document}